\documentclass[aps,pre,preprint,superscriptaddress,showpacs,letter]{revtex4}
\usepackage{amsfonts,amssymb,amsmath,latexsym,epsfig} 
\usepackage{amsmath}
\usepackage{graphicx,epsfig}

\begin{document}

\title{Stochastic Resonance and Dynamic First-Order Pseudo-Phase Transitions in the Irreversible Growth of Thin Films under Spatially Periodic Magnetic Fields} 

\author{Ernesto S. Loscar}
\affiliation{
Instituto de Investigaciones Fisicoqu\'imicas  Te\'oricas y Aplicadas,\\
 (INIFTA), CCT-La Plata CONICET, Facultad de Ciencias Exactas, Universidad Nacional de La Plata, C.C. 16 Suc. 4, 1900 La Plata, Argentina}
 \author{Juli\'an Candia}
 \affiliation{
 Instituto de F\'{\i}sica de L\'{\i}quidos y Sistemas 
 Biol\'ogicos,\\ (CONICET, UNLP), 59 Nro 789, 1900 La Plata, Argentina}
 \affiliation{Department of Physics, 
 University of Maryland, College Park, MD 20742, USA}

\begin{abstract}
We study the irreversible growth of magnetic thin films under the influence of spatially periodic fields by means of extensive Monte Carlo 
simulations. We find first-order pseudo-phase transitions that separate a dynamically disordered phase from a dynamically  
ordered phase. 
By analogy with time-dependent oscillating fields applied to Ising-type models, we qualitatively associate this dynamic transition with 
the localization/delocalization transition of ``spatial hysteresis" loops. Depending on the 
relative width of the magnetic film, $L$, compared to the wavelength of the external field, $\lambda$, different transition regimes are observed.  
For small systems ($L<\lambda$), the transition is associated with the Standard Stochastic Resonance regime, while, for large 
systems ($L>\lambda$), the transition is driven by Anomalous Stochastic Resonance. The origin of the latter 
is identified as due to the emergence of an additional relevant lengthscale, namely the roughness of the spin domain switching interface. 
The distinction between different stochastic resonance regimes is discussed at length, both qualitatively by means of snapshot configurations, as well as 
quantitatively via residence-length and order-parameter probability distributions. 
\end{abstract}         


\maketitle

\section{Introduction}
With the sustained progress of nanoscale deposition techniques such as sputtering and molecular beam epitaxy, which grant control over the deposition process at the atomic scale, thin films are playing an ever increasing role in applied science and technology~\cite{buns94,maha00,ohri02,gloc10}. Besides the use of thin films in many applications of great technological importance such as optical coatings, electronics, packaging, and magnetic recording media, much effort has recently been focused on their investigation under a variety of experimental conditions (see e.g.~\cite{xu98,lu04,jung04,lu05,wang11} and references therein). 

Recently, several experimental investigations have characterized the response of various nanoscale magnetic systems under spatially-varying magnetic fields. 
The physics of microscopically inhomogeneous magnetic fields relates to important fundamental problems in the fractional quantum Hall effect, superconductivity, spintronics, and graphene 
physics (see~\cite{noga10} and references therein). 
Buchholz et al. investigated quantum dots in spatially periodic magnetic fields, where a rich spectral behavior was observed 
for varying parameters (amplitude, wavelength, and phase) of the periodic magnetic field~\cite{buch06}.     
Very recently, double spin resonance in a spatially periodic magnetic field with zero average has been experimentally observed as well~\cite{noga11}. 
Furthermore, Davidenko and Al-Kadhimi studied the formation of magnetic gratings in epitaxial garnet films under spatially 
periodic magnetic fields, which were obtained by using spatially periodic fringing fields from a magnetic tape~\cite{davi04}.

Moreover, a number of theoretical efforts have been devoted to study the response of magnetic systems under inhomogeneous and/or oscillating magnetic fields. 
The interplay of characteristic time- and lengthscales between the geometrical and dynamical features of the magnetic system (shape, size, thickness, mono- or multilayer configuration, deposition rate, etc) and the external magnetic fields leads to a variety of complex phenomena. Some of these studies, which focused on Ising-like spin systems, have 
investigated field-driven dynamical phase transitions~\cite{lo90,side98a,korn02,park13}, dynamical symmetry breaking of hysteresis loops~\cite{chak99, akta12}, 
switching and magnetization reversal~\cite{miya98, liu08, berg12}, droplets, nucleation, and metastable states~\cite{rikv94,korn00,park04,liu10}, and stochastic resonance~\cite{side98,evst05,saho08}.

Within the broad context of these recent experimental and theoretical investigations, 
the aim of this work is to address the computational modeling of far-from-equilibrium 
thin film growth under spatially periodic magnetic fields by means of extensive Monte Carlo simulations. 
The irreversible growth of magnetic thin films is investigated by using the so-called magnetic Eden 
model (MEM)~\cite{ausl93,cand01,cand11}, an extension 
of the classical Eden model~\cite{eden58} in which particles 
have a two-state spin as an additional degree of freedom. 
The MEM growth process is irreversible, since newly deposited particles are not allowed to flip and thermalize once they 
are added to the growing cluster.  

Driven by the external magnetic field, we observe the occurrence of Stochastic Resonance (SR) phenomena leading to a first-order phase transition between a dynamically symmetric phase and a dynamically asymmetric phase. This transition can be associated to 
the phenomenon of ``spatial hysteresis", which is analogous to the behavior of Ising-like spin systems under (time-dependent) 
oscillating magnetic fields. 
However, new features arise from the nature of the growth process investigated here: the roughness of the interface between neighbor 
magnetic domains has a characteristic scale in competition 
with the wavelength of the external magnetic field. Therefore, depending on the 
interplay of characteristic lengthscales, two kinds of stochastic resonance regimes are possible, which we further describe as Standard (SSR) and Anomalous (ASR) Stochastic Resonance. The characteristics of both SSR and ASR regimes are extensively discussed.  

The paper is laid out as follows. In Section 2, we introduce the model and 
describe the Monte Carlo algorithm used to simulate thin film growth under spatially periodic magnetic fields. 
In Section 3, we present and discuss our results. Finally, Section 4 contains our conclusions.

\section{The Model and the Monte Carlo Simulation Method}
In the Eden model~\cite{eden58}, originally proposed as a stochastic kinetic model for the growth of bacterial colonies, particles are added at random to the perimeter of a growing cluster. The Eden model is known to belong to the Kardar-Parisi-Zhang (KPZ)  universality class~\cite{bara95}; indeed, the most accurate simulation results for the KPZ model~\cite{mari00} appear to agree well with some of the formerly reported exponents for KPZ~\cite{amar90} and the Eden model in $(d+1)$-dimensions for $d=1,2$~\cite{devi89}.
The magnetic Eden model (MEM) is an extension of the Eden model in which particles have a magnetic moment coupled through Ising-like interactions. In regular lattices, the MEM's growth process leads to Eden-like self-affine growing interfaces and fractal cluster structures in the bulk and displays a rich variety of nonequilibrium phenomena, such as thermal order-disorder continuous  phase transitions, spontaneous magnetization reversals, as well as morphological, wetting, and corner-wetting transitions 
(see Ref.~\cite{cand08} for a review). 
  
In this work, the MEM is studied in $(1 + 1)-$dimensional strip geometries by using a rectangular substrate of 
size $L\times M$, where $M\gg L$ is the growth direction. 
The location of each spin on the lattice is specified through its 
coordinates $(x,y)$ ($1\leq x\leq M$, $1 \leq y \leq L$). 
The starting seed for the growing cluster is a column of $L$ 
parallel-oriented spins placed at $x=1$ and cluster growth takes place 
along the positive longitudinal direction (i.e., $x\geq  2$). 
We adopt continuous boundary conditions along the $y-$direction, i.e. sites $(x,y=1)$ are nearest-neighbors to sites $(x,y=L)$. 
Since the films are effectively semi-infinite and the substrate length along the 
growth direction plays no role, the only characteristic length of the setup is the transverse linear 
size $L$.    

In this work, we study the irreversible growth of thin films under spatially periodic magnetic fields given by:
\begin{eqnarray}
H(x)=h_0\ {\rm sin}(2\pi x/\lambda)\ .
\label{field}
\end{eqnarray}
\noindent We consider that fluctuations are controlled by a thermal bath that maintains the temperature fixed at $T$. 
According to the MEM's growth rules~\cite{ausl93}, 
clusters are grown by selectively adding two-state spins ($S_{xy}= \pm 1$) to perimeter sites, 
which are defined as the nearest-neighbor (NN) empty sites of the already occupied ones. 
Considering a ferromagnetic interaction of strength $J>0$ between NN spins, 
the energy $E$ of a given configuration of spins is given by 
\begin{eqnarray}
E=-\frac{J}{2}\sum_{{\langle xy,x'y'\rangle}_{NN}}
S_{xy}S_{x'y'}-\sum_{xy} H(x)S_{xy}\  ,
\label{energy}
\end{eqnarray}
\noindent where the first summation is taken over occupied pairs of NN sites, while the second term 
accounts for the interaction between the magnetic field and all deposited spins.  
The Boltzmann constant is set equal to unity throughout;  
temperature, magnetic field, and energy are measured 
in units of $J$. The probability for a perimeter site at $(x,y)$ 
to be occupied by a spin is proportional to the Boltzmann factor
$\exp(- \Delta E/T)$, where $\Delta E$
is the change of energy involved in the addition of
the spin. Notice that the adoption of a Boltzmann factor implies that, near the growing interface, not-yet-deposited spins 
in the spin gas reservoir are in equilibrium with the thermal bath at temperature $T$; upon attachment to the interface, however, 
spins become quenched. Indeed, although Eq.~(\ref{energy}) resembles the Ising Hamiltonian, 
the MEM is a nonequilibrium model in which, as new spins are continuously added, 
older spins remain frozen and are not allowed to flip, detach, nor diffuse.   

At each step, all perimeter sites have to be considered and the probabilities 
of adding a new (either up or down) spin to each site must be evaluated. 
Using the Monte Carlo simulation method, after all probabilities are 
computed and normalized, the growth site and the orientation of the 
new spin are both simultaneously determined by means of a pseudo-random number. 
The change of energy involved in the addition 
of a new spin, $\Delta E$, depends on the local configuration of neighboring spins; however,    
since the sum of Boltzmann factors over all perimeter sites is a global (nonconserved) quantity, the MEM's growth rules require updating the normalized deposition probabilities at 
each time step and lead to very slow algorithms compared with analogous equilibrium spin models.
Clusters having up to $10^9$ spins have typically been grown for lattice sizes up to $L=1024$.
 
As in the case of the classical Eden model~\cite{eden58}, the magnetic Eden model 
leads to a compact bulk and a self-affine growth interface~\cite{ausl93}. 
The growth front may temporarily create voids within the bulk, usually not far from the rough growth interface. 
However, since the boundaries of these voids are also perimeter sites, they ultimately 
become filled at some point during the growth process. 
Hence, far behind the active growth interface, the system is compact 
and frozen, and the different quantities of interest can thus be measured on defect-free transverse columns.  
The growth of magnetic Eden aggregates in $(1+1)$-strip geometries is 
characterized by an initial transient 
length $\ell_{tr}\sim L$ (measured along the growth direction, i.e. the $x-$axis) 
followed by a nonequilibrium stationary state that is independent of the initial 
configuration~\cite{cand01}. We considered starting seeds formed by $L$ up spins (i.e. $S_{(x=1,y)}=1$) 
but any choice for the seed leads to the same stationary states for $x\gg\ell_{tr}$.  
By disregarding the transient region, all results reported in this paper are 
obtained under stationary conditions.
While keeping the field's period fixed at $\lambda=100$ lattice units throughout, 
we explore extensively the remaining parameter space by scanning meaningful ranges in temperature, magnetic field amplitude, and system size. 
Notice that, since the transitions are due to the competition between $\lambda$ and 
the other relevant lengthscales in the system, assuming a fixed value for $\lambda$ throughout does not entail a loss of 
generality. 

\section{Results and Discussion}
\subsection{Small Systems: the Standard Stochastic Resonance Regime.}
\begin{figure}[pt]
\begin{center}
\epsfxsize=6.4truein\epsfysize=3.8truein\epsffile{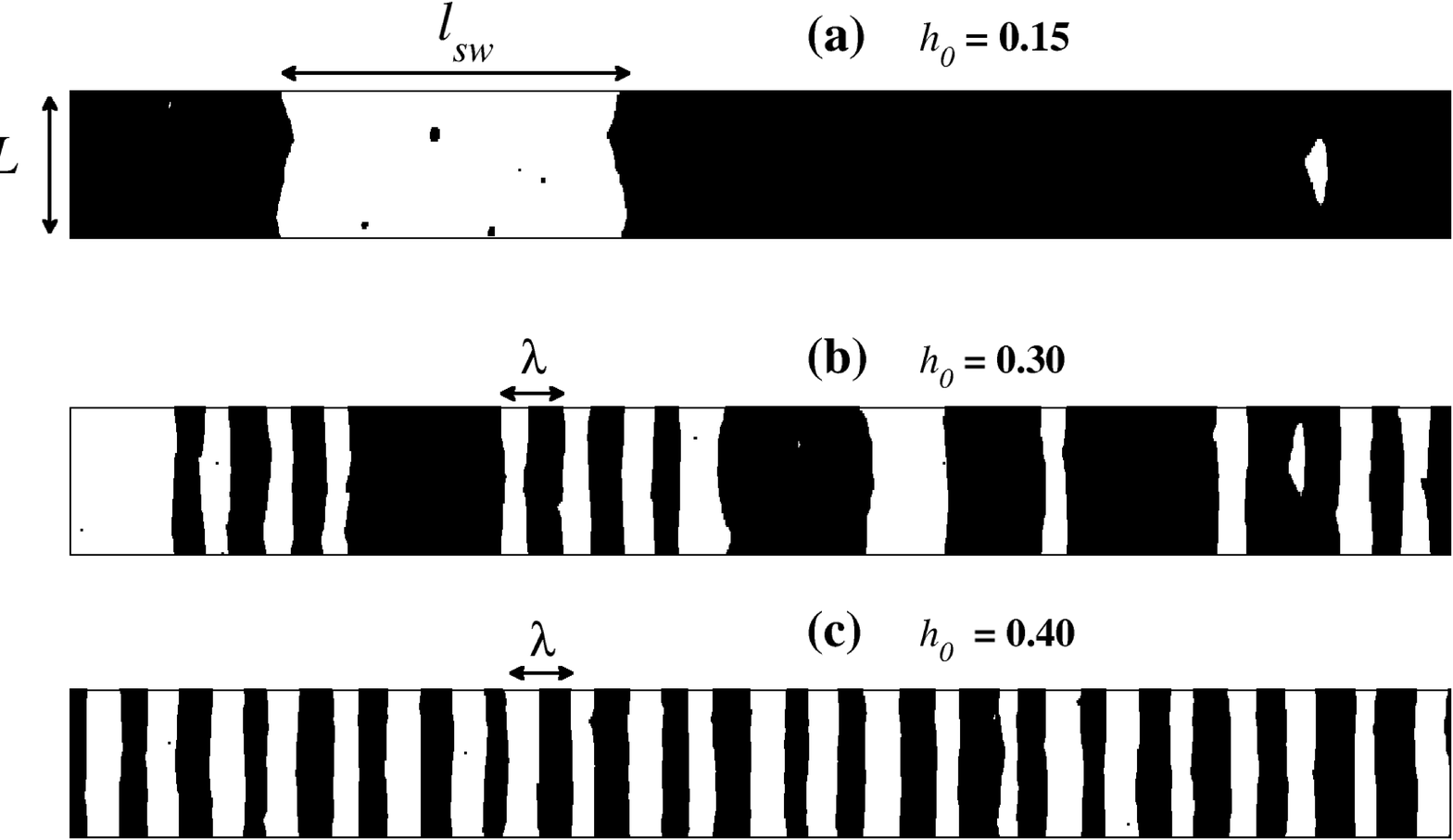}
\caption{Snapshots of characteristic growth regimes for thin films with a small system size, $L=64$, growing at temperature $T=0.3$ 
under a magnetic field of period $\lambda=100$ and different amplitudes: (a) $h_0=0.15$, 
(b) $h_0=0.30$, and (c) $h_0=0.40$. The length of one domain between consecutive magnetization switching events, $\ell_{sw}$, 
is indicated on the top panel.}
\label{fig:snap64}
\end{center}
\end{figure}

In order to gain qualitative insight, let us first investigate some typical modes of 
thin film growth, as displayed by the snapshots of Fig.~\ref{fig:snap64} for different field amplitudes: (a) $h_0=0.15$, 
(b) $h_0=0.30$, and (c) $h_0=0.40$. These snapshots correspond to thin films of size $L=64$ and temperature $T=0.30$. 
This system size is small compared to the magnetic field's period ($\lambda=100$). 
In the next Subsection, we will discuss the 
case of large systems with $L>\lambda$, in which new phenomena emerge from the competition of characteristic lengthscales.  
Since we are concerned with compact and frozen thin films in the stationary regime,  
the seed and seed-dependent transient region have been discarded. 
In Fig.~\ref{fig:snap64}, the active growth interface (to which new spins are attached 
during the film's growth process) is not shown; the snapshots correspond to fully grown regions spanning a length of 23 field periods.

In Fig.~\ref{fig:snap64}(a), the bulk grows ordered and the field favors small fluctuations: the system grows with most spins 
aligned in the same direction, while the field drives the formation of small clusters of opposing magnetization. 
The small field amplitude, however, is not capable of fully reversing the bulk magnetization except for magnetization 
switching events, which occur only sporadically over very long lengthscales $\ell_{sw}\gg\lambda$. 
The period-averaged magnetization, 
\begin{eqnarray}
Q=\langle{\ \frac{1}{\lambda}\int_{x_0}^{x_0+\lambda}m(x){\rm{d}}x}\ \rangle_{x_0}\ ,
\label{OP_eq}
\end{eqnarray}
\noindent is close to $Q\sim \pm 1$. The brackets in Eq.(\ref{OP_eq}), $\langle ... \rangle_{x_0}$, 
denote averages taken over periods $\lambda$ within the region where the film is fully grown, where 
$x_0$ are multiples of $\lambda$ and $m(x)$ is the normalized magnetization of the column of spins at position $x$.   
In Fig.~\ref{fig:snap64}(b), the field is strong enough to drive the system through frequent magnetization switching events, 
leading to the formation of ordered transverse strips whose magnetization alternates between 
the up and down directions. These bands have uneven widths, but they all roughly correspond to odd multiples of the half-period $\lambda/2$. 
Averaging the magnetization over one period as in Eq.~(\ref{OP_eq}), the longer, well-ordered domains contribute to $Q\sim \pm 1$, 
while the shorter, up/down domain sequences contribute to $Q\sim 0$. 
Indeed, the snapshot in Fig.~\ref{fig:snap64}(b) marks the occurrence of a pseudo-phase transition driven by 
stochastic resonance with the external magnetic field, 
a phenomenon that will be further characterized below.  
Fig.~\ref{fig:snap64}(c) shows a growth mode where the magnetic field is strong enough to fully reverse the bulk 
magnetization within each cycle, therefore forming ordered transverse strips whose magnetization alternates between 
the up and down directions at regular intervals. Since the width 
of these strips is roughly equal to $\lambda/2$, the mean magnetization averaged over a cycle is close to $Q\sim 0$. 
It is worth noticing that the width of the spin domain switching interface, $W_{sw}$, i.e. the roughness of the interface 
formed between consecutive up and down spin domains, is much smaller than the field period $\lambda$, as expected from 
the fact that the system size $L$ is small compared to $\lambda$. 
In the next Subsection, when we discuss larger systems 
with $L>\lambda$, we will find additional effects arising from the interplay of $W_{sw}$ with the other characteristic lengthscales of the system. 

\begin{figure}[pt]
\begin{center}
\epsfxsize=5.5truein\epsfysize=4truein\epsffile{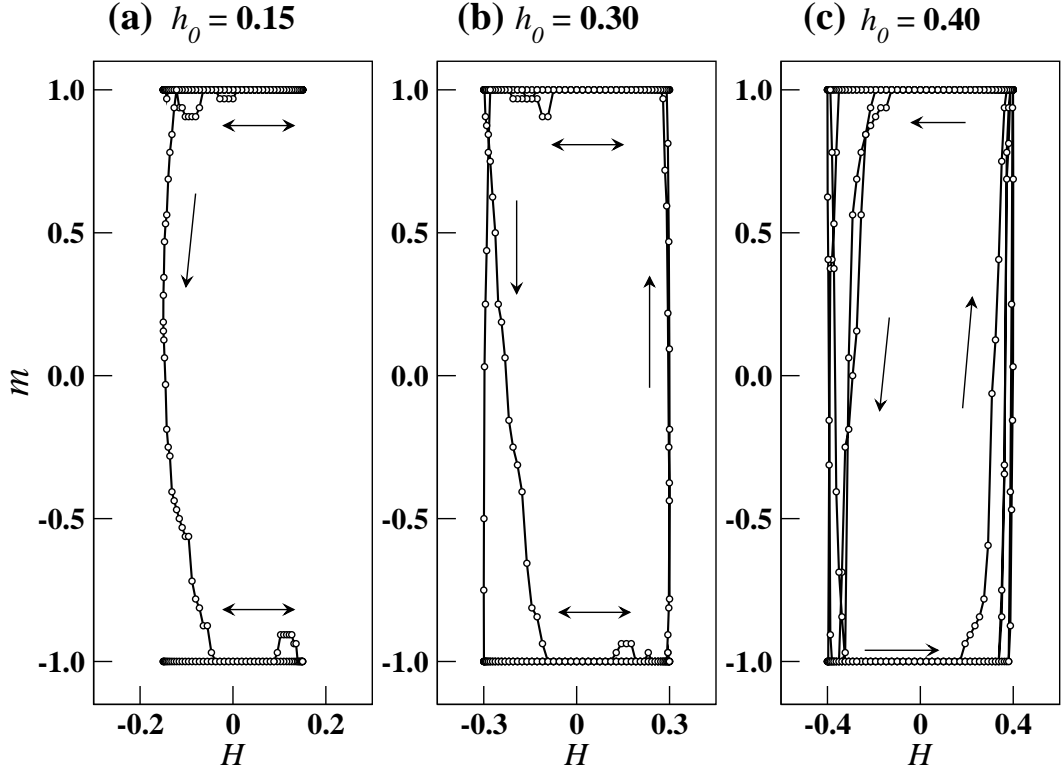}
\caption{{\it Spatial hysteresis} loops using data from the snapshots of Fig.~\ref{fig:snap64}. 
The magnetic field amplitudes are (a) $h_0=0.15$, 
(b) $h_0=0.30$, and (c) $h_0=0.40$, as indicated.}
\label{loops}
\end{center}
\end{figure}

Let us now explore the trajectories on the magnetization-vs-field plane, i.e. $m(x)$ as a function of $H(x)$,
using data from the snapshots of Fig.~\ref{fig:snap64}. 
Interestingly, these 
trajectories can be regarded as {\it spatial hysteresis} loops, i.e. a phenomenon akin to the more usual (time) hysteresis loops 
observed in spin systems under (time-dependent) oscillating fields. Indeed, when the field amplitude is small (Fig.~\ref{loops}(a)), 
the loops (indicated by horizontal arrows) remain pinned 
to the $m\sim\pm 1$ ordered regions and exhibit a negligible area. Over a lengthscale $\ell_{sw}\gg\lambda$, a passage from the $m\sim 1$ region 
to the $m\sim -1$ region is observed (down arrow). By increasing $h_0$, a transition to a mixed state is observed (Fig.~\ref{loops}(b)), 
where small-area loops pinned to $m\sim\pm 1$ (horizontal arrows) alternate with large-area loops (vertical arrows) due to 
the periodic (but out-of-phase) response of the magnetization to the applied field. As $h_0$ is further increased, the small-area loops 
vanish and the large-area loops prevail, as shown in Fig.~\ref{loops}(c). 

By exploiting the analogy with spin systems under oscillating fields, such as e.g. the kinetic Ising model in sinusoidally oscillating 
magnetic fields~\cite{lo90}, we can gain some insight into the novel phenomenon of spatial hysteresis. 
If the periodic field (Eq.(\ref{field})) were replaced by a ``step function field" that would switch suddenly, at some position $x^*$, from $h_0$ 
to $-h_0$, the mean column magnetization would drop from $m\sim 1$ to $m\sim -1$ over a characteristic 
``magnetization decay length" $\ell_D$. Naturally, this characteristic length would depend on the field strength and would become shorter as the 
field amplitude is increased. The far-from-equilibrium response under spatially periodic fields can therefore 
be viewed as a competition between two lengthscales, namely the characteristic decay length $\ell_D$ versus the 
magnetic field's period $\lambda$.
For sufficiently large field amplitudes, $\ell_D\ll\lambda$ and, thus, the field is capable of switching the film's magnetization within the 
length of one field half-period. This leads to the formation of ordered strips with $\ell_{sw}\sim\lambda/2$ (Fig.~\ref{fig:snap64}(c)),  
which corresponds to the (symmetric) disordered dynamic phase with $Q\sim 0$. 
For sufficiently small field amplitudes, on the other hand, $\ell_D\gg\lambda$ and, therefore, the magnetization does not switch within 
one field half-period. Rather, the film keeps growing with all spins mostly 
parallel-aligned (Fig.~\ref{fig:snap64}(a)) and magnetization switching events occur over much longer lengthscales, $\ell_{sw}\gg\lambda$, 
which correspond to the (asymmetric) ordered dynamic phase with $Q\sim \pm1$. 
From Fig.~\ref{loops}, we observe that this dynamic phase transition is associated with a localization/delocalization transition 
of spatial hysteresis loops, 
where symmetry breaking is driven by the amplitude of the spatially-periodic magnetic field. 

\begin{figure}[pt]
\begin{center}
\epsfxsize=3truein\epsfysize=5.5truein\epsffile{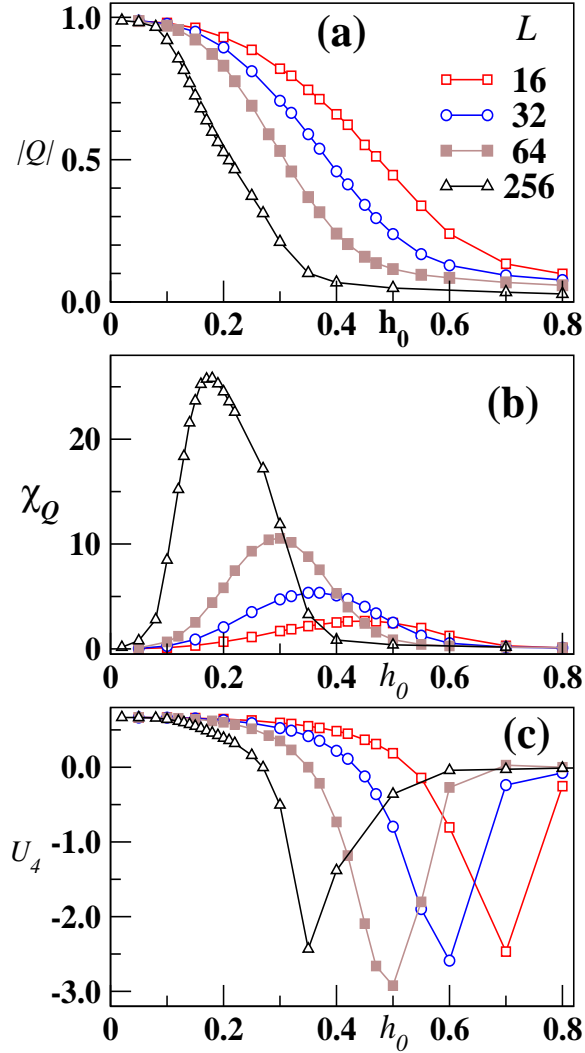}
\caption{(Color online) Simulation results of thin film growth under spatially periodic magnetic fields 
as a function of the field amplitude $h_0$, for $T=0.30$, $\lambda=100$, and 
different film sizes, as indicated.
(a) Absolute value of the period-averaged magnetization, $|Q|$. (b)
Susceptibility, $\chi_Q$. (c) Fourth-order cumulant of the order parameter probability distribution, $U_4$.
Film sizes are represented by the same symbols in panels~(a)-(c).}
\label{OP}
\end{center}
\end{figure}

The natural order parameter to study dynamic phase transitions (DPTs) is the period-averaged magnetization given by Eq.(\ref{OP_eq}). 
Due to finite-size effects, however, we will use instead the absolute value of the order parameter, $|Q|$, which prevents spurious 
averaging to zero (since, even in the $h_0\to 0$ limit, the system is capable of switching from $Q\sim 1$ to 
$Q\sim -1$, or viceversa, due to large thermal fluctuations at scales comparable to the system size). 
Fig.~\ref{OP}(a) shows $|Q|$ as a function of the field amplitude $h_0$ for $T=0.30$ and different system sizes, as indicated. 
In agreement with the previous discussion, the system undergoes a field-driven transition from the asymmetric dynamic phase ($|Q|\sim 1$) 
for small field amplitudes to the symmetric dynamic phase ($|Q|\sim 0$) for large field amplitudes. Rigorously, this transition should be 
regarded as a {\it pseudo}-phase transition that affects the growth of thin magnetic films of finite size. As shown by Fig.~\ref{OP}(a), larger system 
sizes cause the plots to shift to smaller field amplitudes. 
It should be remarked that the original MEM in $(1+1)$ dimensions is non-critical (i.e. the finite-size order-disorder critical temperature $T_c(L)$ tends to zero as $L\to\infty$~\cite{cand01}) and therefore there is no bulk ordered phase.
Further discussions on finite-size effects will be presented below.

For equilibrium systems, the magnetic susceptibility is related to order parameter fluctuations 
by the fluctuation-dissipation theorem. Although the validity of a fluctuation-dissipation relation in the case of nonequilibrium systems
is not formally proven, earlier studies of nonequilibrium spin models~\cite{side98a,korn00} have shown that 
assuming an analogous definition for the susceptibility in terms of the moments of the order parameter probability distribution, namely 
\begin{equation}
\chi_Q = \frac{L}{T}\left(\langle Q^2\rangle-\langle|Q|\rangle^2\right)\ ,
\label{chi_eq}
\end{equation}
is useful to investigate the nature of phase transition phenomena under far-from-equilibrium conditions. 

Fig.~\ref{OP}(b) shows plots of $\chi_Q$ as a function of the field amplitude $h_0$ for $T=0.30$ and different system sizes, as indicated. Similarly to the behavior of the order parameter in Fig.~\ref{OP}(a),  
the peaks of the susceptibility appear 
sharper and shifted to smaller field amplitudes as the system size is increased. 

The Binder cumulant, defined by
\begin{eqnarray}
U_4=1-\frac{\langle Q^4\rangle}{3\langle Q^2\rangle^2}\  ,
\label{binder_eq}
\end{eqnarray}
\noindent is a fourth-order cumulant dependent on the variance and the kurtosis of the order parameter probability 
distribution. Since, for second-order phase transitions, the scaling prefactor of the cumulant is independent of the 
sample size, plots of $U_4$ versus the control parameter 
lead to a common (size-independent) intersection point that corresponds to the location of the critical 
value of the order parameter in the thermodynamic limit~\cite{bind81}. In contrast, for first-order phase transitions, 
a characteristic signature of $U_4$ is a sharp fall towards negative values~\cite{bind84,chal86}. 

Fig.~\ref{OP}(c) shows plots of $U_4$ as a function of the field amplitude $h_0$ for $T=0.30$ and different system sizes, as indicated. 
These plots display the hallmark behavior for first-order phase transitions, namely a sharp drop towards negative values at the transition point. 
Consistent with the behavior observed above (Figs.~\ref{OP}(a)-(b)), the location of the transition is shifted 
towards smaller values of $h_0$ as $L$ is increased. 
We conclude, therefore, that finite-size films growing irreversibly under spatially-periodic magnetic fields undergo a dynamic pseudo-phase 
transition between an (asymmetric) ordered dynamic phase ($|Q|\sim 1$ for small $h_0$) and a (symmetric) disordered 
dynamic phase ($|Q|\sim 0$ for large $h_0$), whose nature is discontinuous (first-order). 

Besides the minima of the cumulant at the transition points, another characteristic signature of first-order phase transitions is the linear 
scaling behavior of the maxima of $\chi_Q$ (i.e. the height of the peaks shown in Fig.~\ref{OP}(b), which we denote by $\chi_{Q}^{max}$) 
as a function of the volume $L^{D}$, where $D$ is the system's effective Euclidean dimension. 
In contrast, second-order phase transitions are typically characterized by non-linear scaling relations, from which the critical exponent ratio $\gamma/\nu$ 
can be determined~\cite{priv90}.
 
\begin{figure}[pt]
\begin{center}
\epsfxsize=4.5truein\epsfysize=3.5truein\epsffile{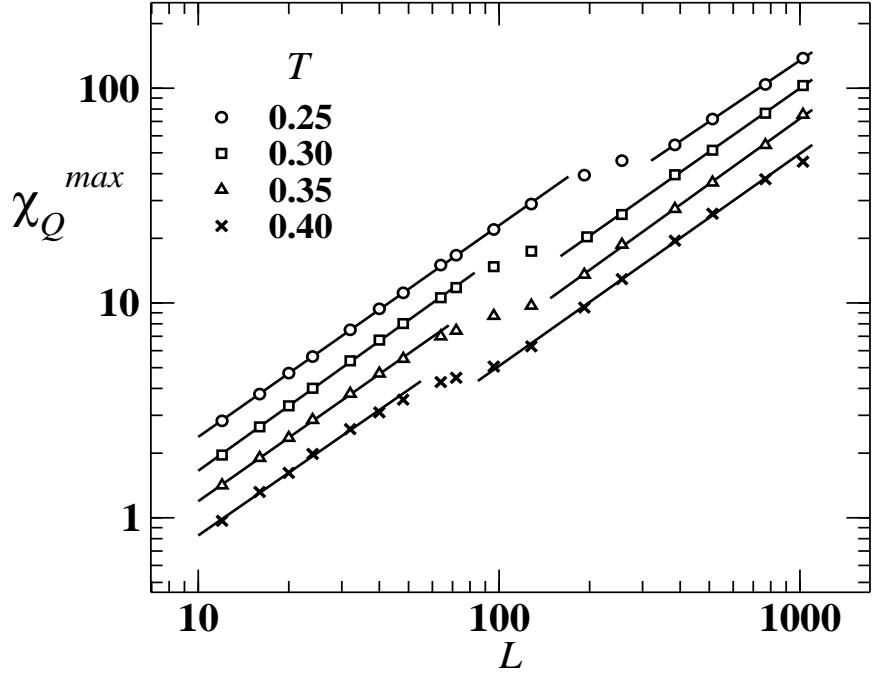}
\caption{Log-log plot of the maxima of $\chi_Q$ as a function of the system size $L$ for different temperatures, as indicated. 
The solid lines show the best fits to the data on the left- and right-hand linear regimes, which are separated by a small crossover region. Within the fitting error bars, all fitted lines are consistent with a slope equal to unity. } 
\label{chimax}
\end{center}
\end{figure}

Fig.~\ref{chimax} shows log-log plots of the maxima of $\chi_Q$ as a function of the system size $L$ for different 
temperatures, as indicated. 
Two separate linear regimes are observed, namely the small-$L$ region with slope equal to $1.00\pm 0.02$ for $L\lesssim 60-100$, and the large-$L$ region 
with slope equal to $0.99\pm 0.02$ for $L\gtrsim 100-150$. Therefore, both regimes are consistent with the linear behavior $\chi_{Q}^{max}\propto L$, 
which is the expected behavior for first-order pseudo-phase transitions in one-dimensional systems.
Intriguingly, however, these two regimes are 
separated by a crossover region that takes place roughly around $L\sim 100$, i.e. at system sizes comparable to the magnetic field's 
period $\lambda=100$. In the remainder of this Subsection, we investigate further the thin film growth regime for small systems,
while the following Subsection will be dedicated to study the growth regime for large systems. Finally, the last Subsection will provide a qualitative explanation for the nature of the crossover behavior.  

\begin{figure}[pt]
\begin{center}
\epsfxsize=6truein\epsfysize=5truein\epsffile{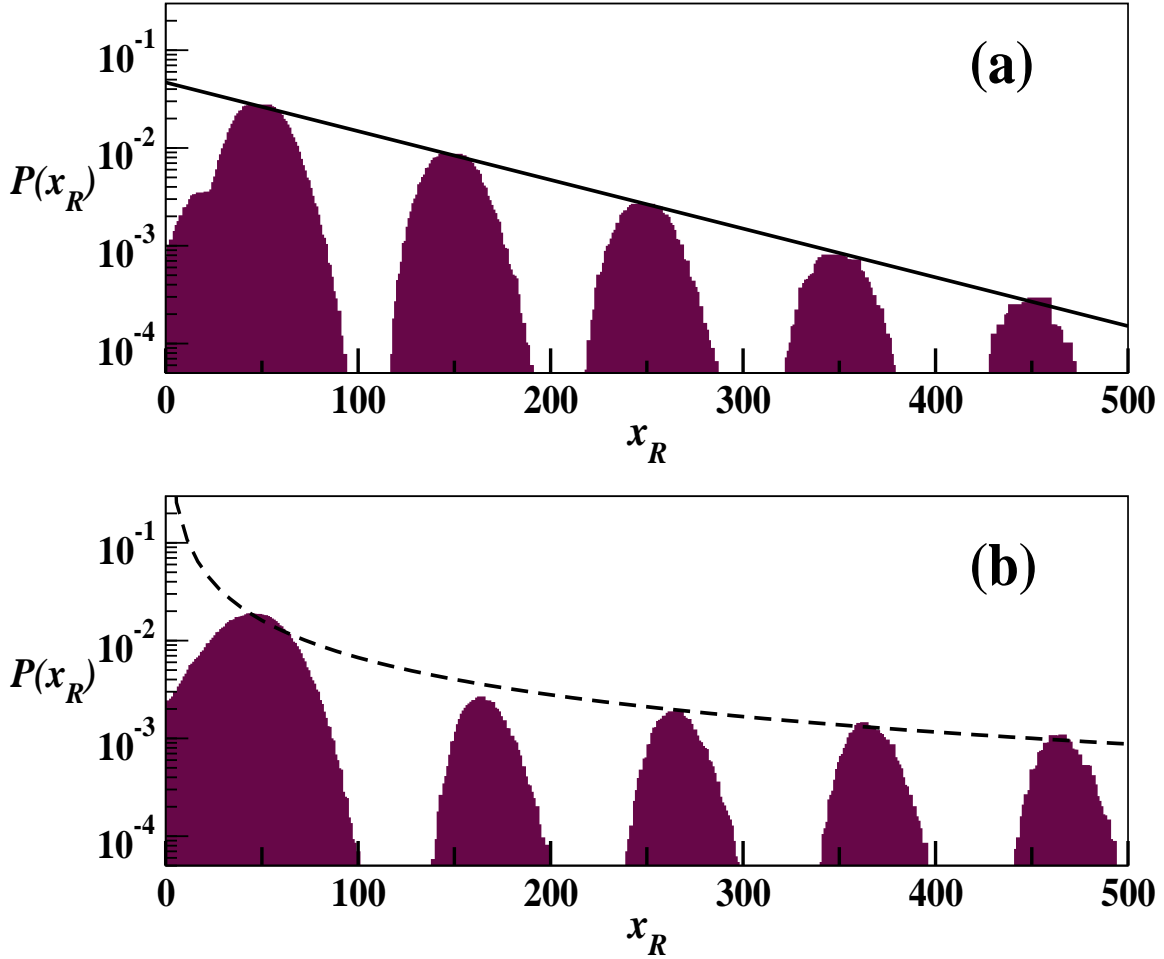}
\caption{(Color online) Log-linear residence-length probability distributions at the dynamic phase transition for $T=0.30$: 
(a) $L=64$ and $h_0=0.30$; (b) $L=256$ and $h_0=0.175$. In (a), the solid line is an exponential 
fit to the distribution peaks given by $P= Ae^{-x_R/B}$, with $A=0.046\pm0.001$ and $B=8.7\pm0.1$. In (b), the dashed line 
is a power-law fit given by $P= C x_R^{-D} $ with $C=2.3\pm 0.5$ and $D=1.26\pm 0.03$.}
\label{residence}
\end{center}
\end{figure}

The well-known phenomenon of stochastic resonance occurs when a system's characteristic scale is matched by the 
characteristic scale of an external field~\cite{gamm98}. In the case of
symmetric bistable systems, this transient regime is revealed by a characteristic exponential 
decrease of the maxima of the residence-time probability distributions. Moreover, the residence-time peaks appear at regular 
locations that correspond to odd multiples of $\lambda/2$, thus indicating that resonance with the external field is indeed 
the mechanism responsible for driving the system across different dynamic states. 
 
Fig.~\ref{residence} shows the log-linear residence-length probability distribution at the dynamic phase transition for $T=0.30$, $L=64$, and $h_0=0.30$. Analogously to the definition of 
resident time in bistable systems under time-dependent oscillating fields~\cite{korn02}, ``residence length" 
is here defined as the length (measured along the longitudinal growth direction) 
between two consecutive crossings of the column-averaged 
magnetization profiles across $m=0$. I.e., if $x_0$ is the position of one magnetization switch (e.g., from $m(x_0-1)\geq 0$ 
to $m(x_0+1)<0$) and $x_1$ is the position for the next magnetization switch (correspondingly, from $m(x_1-1)<0$ 
to $m(x_1+1)\geq 0$)), the residence length is computed as 
$x_R=x_1-x_0$. By growing very long magnetic films (for which many such magnetization switching events are observed), we 
obtain residence-length probability distributions. 
Fig.~\ref{residence}(a) shows the behavior expected for a standard stochastic resonance regime: the distribution peaks are located at 
odd multiples of the half-period $\lambda/2$ and their heights decrease exponentially, as evidenced by the 
straight solid line in the log-linear plot, which is a fit to $P= Ae^{-x_R/B}$ with $A=0.046\pm0.001$ and $B=8.7\pm0.1$. 
Based on these findings, 
we will refer to the region for small lattice sizes compared to the field period as the ``Standard Stochastic Resonance" (SSR) regime. 
This corresponds to the linear regime observed on the 
left-hand side of Fig.~\ref{chimax}.

\subsection{Large Systems: the Anomalous Stochastic Resonance Regime.}

For magnetic film sizes larger than the field period, a different growth regime appears. 
This is indeed apparent from the resident-length probability distribution shown in Fig.~\ref{residence}(b), 
where the distribution peaks appear shifted from the expected locations at odd multiples of $\lambda/2$ and, moreover, the maxima depart from the expected exponential  
relation. The best fit to the local maxima is found by using a power-law relation $P= C x_R^{-D} $, with $C=2.3\pm 0.5$ and $D=1.26\pm 0.03$, shown by the dashed line in Fig.~\ref{residence}(b). 

\begin{figure}[pt]
\begin{center}
\epsfxsize=6.48truein\epsfysize=2.8truein\epsffile{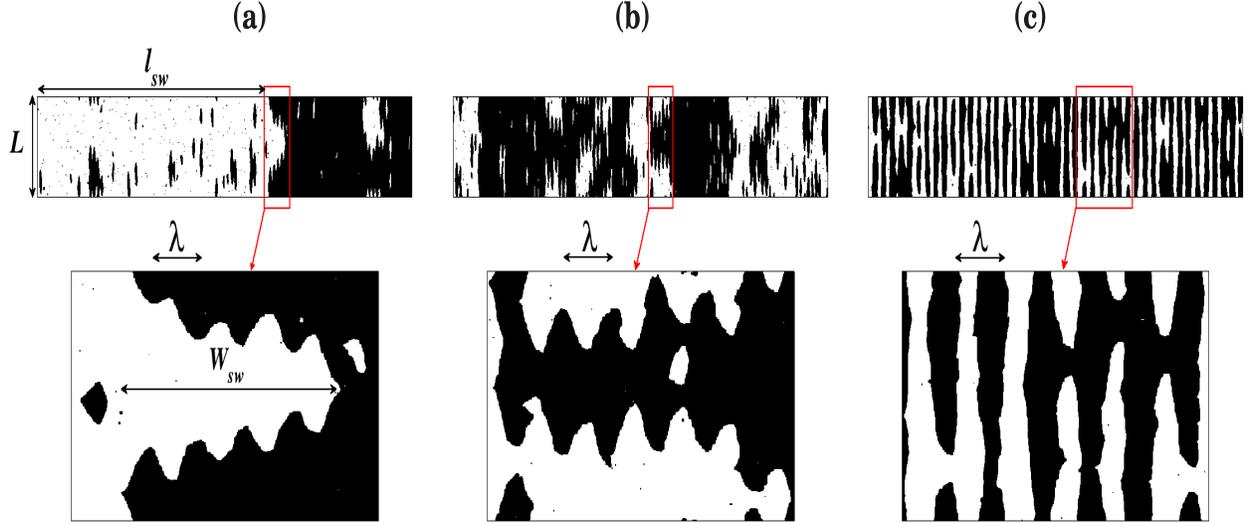}
\caption{(Color online) Snapshots of characteristic growth regimes for a large system size, $L=256$, growing at temperature $T=0.3$ 
under a magnetic field of period $\lambda=100$ and different amplitudes: (a) $h_0=0.12$, 
(b) $h_0=0.175$, and (c) $h_0=0.30$. The characteristic lengthscales are indicated. See more details in the text.}
\label{fig:snap256}
\end{center}
\end{figure}

In order to gain further insight into this growth regime, Fig.~\ref{fig:snap256} shows snapshots for a film size larger than 
the field period (namely, $L=256$ and $\lambda=100$), 
for $T=0.3$ and different field amplitudes: (a) $h_0=0.12$, 
(b) $h_0=0.175$, and (c) $h_0=0.30$. The regions marked by red boxes on the top panels appear expanded in 
the bottom panels, so as to zoom into the regions where spin domain switching events take place. 
Figure~\ref{fig:snap256}(a) shows that, for small field amplitudes, the magnetization switching events occur over lengthscales $\ell_{sw}\gg\lambda$, 
which corresponds to the ordered dynamic phase 
with $|Q|\sim 1$, similarly to Fig.~\ref{fig:snap64}(a). 
However, by zooming into the spin domain switching region (marked with a box and expanded in the bottom panel), we observe that 
spin domain switching takes place over several wavelengths of the periodic magnetic field. 
Following the oscillations of the magnetic field, the spin domain interface displays a wavy pattern and has a width 
$W_{sw}\gg\lambda$. 
Figure~\ref{fig:snap256}(b) corresponds to the phase transition, where the interplay of characteristic 
lengthscales leads to a complex wavy pattern. In this case, distinct transverse bands are observed; however, rather than establishing 
well-separated domains, these transverse strips appear joined with the neighboring domains due to the competition of lengthscales at the transition, 
i.e. $W_{sw}\sim\lambda$. 
Only by increasing the field amplitude well beyond the transition point, as shown in Fig.~\ref{fig:snap256}(c), 
the magnetic field is able to enforce sharp spin domain switching events, thus restoring the lengthscale relation $W_{sw}<\lambda$ and,  consequently, lead to the dynamic disordered phase with $|Q|\sim 0$, similarly to Fig.~\ref{fig:snap64}(c).
We characterize the growth regime for large film sizes (which corresponds to the right-hand side region of Fig.~\ref{chimax}) as the ``Anomalous Stochastic Resonance" (ASR) regime. 
 
\subsection{Discussion}
 
The key to understanding the crossover from the SSR to the ASR regime is to notice that, analogously to other kinds 
of surface growth phenomena, thin films grow 
with a rough interface, whose saturated width in the stationary regime scales with the lattice size as $w\propto L^\alpha$~\cite{vics92,bara95}. For magnetic Eden model thin films, $w= a\times L$ with $a<1$~\cite{cand12}.   
The snapshots from Fig.~\ref{fig:snap64} correspond to a lattice size ($L=64$) smaller than the field period ($\lambda=100$). Therefore,  
the width of the domain switching interface, $W_{sw}$, i.e. the roughness of the interface 
formed between consecutive up and down spin domains (which is of the same order as the width of the growing interface) 
is smaller than the field period $\lambda$. This is the main characteristic of the transition driven by the SSR regime. 
On the other hand, 
when the width of the domain switching interface is comparable to (or larger than) the field 
wavelength, as shown in Fig.~\ref{fig:snap256}, the phase transition is driven by the ASR regime. 
Therefore, the SSR/ASR crossover is due to the competition of characteristic lengthscales, namely the spin domain width 
interface $W_{sw}$ (which depends on $L$ and $h_0$) versus the field period $\lambda$ (which is fixed at $\lambda=100$ lattice units 
throughout). 

\begin{figure}
\begin{center}
\includegraphics[width=8.5cm]{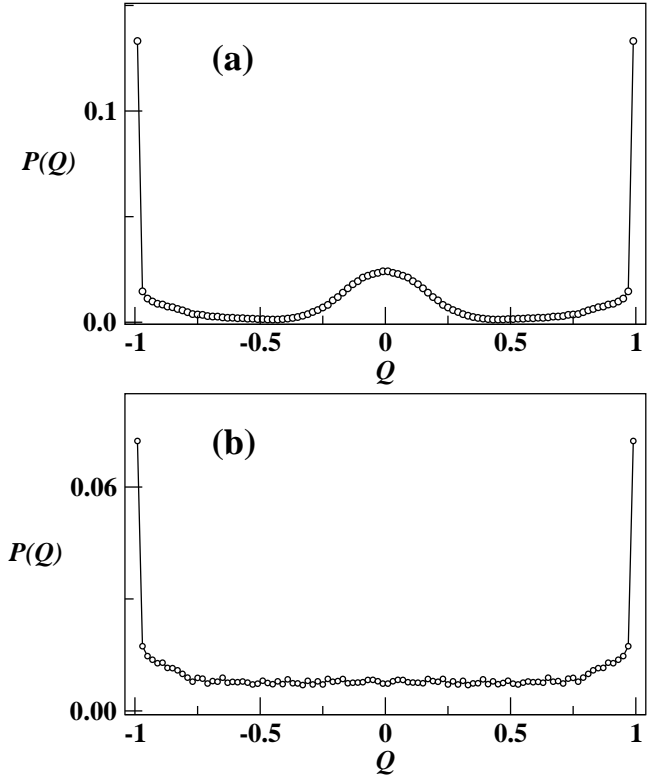}
\caption{Probability distribution functions of the order parameter $Q$ at the transition point for different system sizes: 
(a) For a small system ($L=64$ and $h_0=0.30$), the distribution shows three peaks 
at $Q=\pm 1$ and $Q=0$, characteristic of the Standard Stochastic Resonance (SSR) regime. 
(b) For a large system ($L=256$ and $h_0=0.175$), the distribution shows only 
two peaks at $Q=\pm 1$, while intermediate values have a roughly constant, 
non-zero probability of occurrence. This behavior corresponds 
to the Anomalous Stochastic Resonance (ASR) regime.} 
\label{fig:distribuciones}
\end{center}
\end{figure}

Revisiting Fig.~\ref{chimax}, we observe that the crossover from SSR to ASR occurs for larger system sizes as the temperature 
is decreased. This phenomenon can be qualitatively explained from the fact that, by keeping the system size $L$ fixed, 
the dynamic pseudo-phase transition takes place at larger field amplitudes as the temperature is decreased (i.e. a larger field amplitude is needed to drive the magnetization reversal of an inherently more ordered bulk due to decreased thermal fluctuations). But, since the spin domain width interface becomes smaller for a larger field (because a 
stronger field drives the magnetization switching more sharply), we conclude that, as the temperature is decreased, the spin domain 
width is also decreased (at fixed $L$). On the other hand, at fixed temperature, the spin domain width interface increases with $L$. 
The SSR/ASR crossover occurs for $W_{sw}\sim\lambda$; hence, as the temperature is decreased, we expect the crossover to take 
place at larger system sizes, which agrees with the results from Fig.~\ref{chimax}.  

\begin{figure}
\begin{center}
\includegraphics[width=12.5cm]{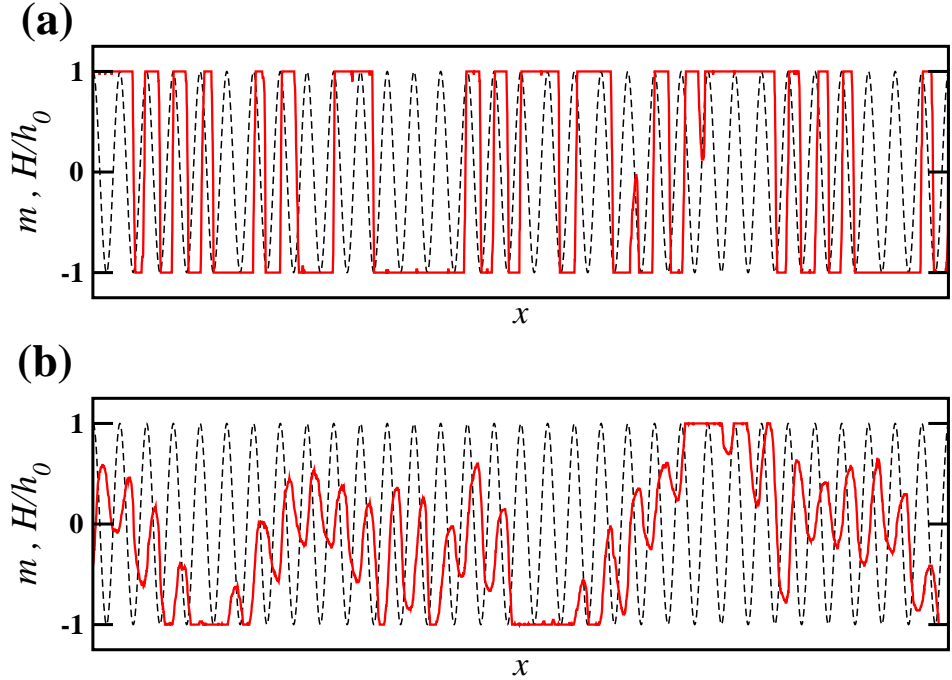}
\caption{(Color online) Magnetization profiles at the transition point for different system sizes (solid red lines) and 
magnetic field $H$ normalized by its amplitude $h_0$ (sinusoidal dashed lines).
(a) For a small system ($L=64$ and $h_0=0.30$), a full magnetization switch can occur within a length of $\lambda/2$, $3\lambda/2$, etc. 
This behavior is characteristic of the SSR regime. 
(b) For a large system ($L=256$ and $h_0=0.175$), the magnetization responds to the field's oscillations at every field cycle, 
but the excursions are only partial. This phenomenon is due to 
the interplay of lengthscales between the domain interface roughness, $W_{sw}$, and the field wavelength, $\lambda$, and gives rise to the ASR regime.}
\label{fig:magprof}
\end{center}
\end{figure}

Let us now consider the probability distribution functions of the order parameter, $P(Q)$, at the transition point. 
Figure~\ref{fig:distribuciones}(a) shows the distribution for $L=64$ and $h_0=0.30$ (which corresponds to the snapshot in 
Fig.~\ref{fig:snap64}(b)), where three peaks are observed at $Q=\pm 1$ and $Q=0$. 
This is, indeed, the expected behavior of a standard stochastic resonance: the system either stays at the dynamically 
ordered phase $Q=\pm 1$ or else, by resonating with the field, the system switches magnetization, thus contributing to the dynamically 
disordered phase $Q=0$. 
The anomalous nature of the ASR regime is shown in Fig.~\ref{fig:distribuciones}(b) for $L=256$ and $h_0=0.175$ (which corresponds to the snapshot in 
Fig.~\ref{fig:snap256}(b)). In this case, due to the interplay of 
lengthscales discussed above, the field is not able to drive the system through a sequence of fully separated 
spin domains. Rather, domains with a mix of up and down spins in a continuum of different proportions are able to form with 
approximately equal probability, as indicated by the flat bottom $P(Q)\simeq const$ for $-0.75\lesssim Q\lesssim 0.75$. Hence, we find that field-resonant stochasticity operates 
in two different modes. In the SSR regime, the stochastic nature is reflected in the length needed to achieve a full magnetization switch, which 
occurs at odd multiples of half-periods $\lambda/2$, $3\lambda/2$, etc. This growth mode is clearly shown by the magnetization profile of Fig.~\ref{fig:magprof}(a). 
In the ASR regime, partial excursions are strongly correlated with 
the magnetic field oscillations; however, the extent to which the magnetization switches is stochastically driven by the field resonance, 
as shown by Fig.~\ref{fig:magprof}(b). Thus, the stochasticity appears to operate along the (longitudinal) growth direction in the SSR regime, 
whereas it appears to operate along the transverse direction in the ASR regime.  

\section{Conclusions}
In this work, we studied the irreversible growth of thin films under spatially periodic magnetic fields in $(1+1)-$dimensional strip geometries. 
By analyzing snapshot configurations, we found qualitative evidence for the occurrence of pseudo-phase transitions from a dynamically 
ordered phase to a dynamically disordered phase, driven by the magnetic field amplitude. By analogy with Ising-like models in time-dependent oscillating fields, 
we characterized the phenomenon of ``spatial hysteresis" and showed evidence of corresponding localization/delocalization transition of spatial hysteresis loops. 
By using the period-averaged magnetization as order parameter, we quantitatively described the dynamic transition. The study of Binder cumulants and the 
susceptibility provided robust evidence of this transition being of discontinuous (first-order) nature.  

Remarkably, two distinct stochastic resonance regimes are responsible for the occurrence of dynamic pseudo-phase transitions, 
depending on the relative width of the thin film compared to the magnetic field's wavelength. For small system sizes, the transition is 
associated with a Standard Stochastic Resonance (SSR) regime. Instead, for large system sizes, the emergence of an additional relevant lengthscale, 
namely the width of the spin domain switching interface, was found to lead to a so-called Anomalous Stochastic Resonance (ASR) regime. 
We analyzed these two stochastic resonance regimes both qualitatively (by means of snapshot configurations) as well as 
quantitatively, via residence-length and order-parameter probability distributions. 

In the context of a great experimental and theoretical interest in magnetic systems growing under inhomogeneous and oscillating magnetic fields, as well as a wide variety of technological applications that benefit from these efforts, we hope that this work will contribute to the progress of this research field and stimulate further work.
   
{\bf  ACKNOWLEDGMENTS}. This work was financially supported by 
CONICET, UNLP, and  ANPCyT (Argentina).

\end{document}